\documentclass[aps,prl,preprint,superscriptaddress]{revtex4-1}

\usepackage{graphicx}

\begin{document}

\renewcommand{\figurename}{Figure}

\title{Fractionalized Wave Packets from an Artificial Tomonaga-Luttinger Liquid}

\author{H. Kamata}
\email[]{kamata.h.aa@m.titech.ac.jp}
\affiliation{Department of Physics, Tokyo Institute of Technology, 2-12-1 Ookayama, Meguro, Tokyo 152-8551, Japan}
\author{N. Kumada}
\affiliation{NTT Basic Research Laboratories, NTT Corporation, 3-1 Morinosato-Wakamiya, Atsugi, Kanagawa 243-0198, Japan}
\author{M. Hashisaka}
\affiliation{Department of Physics, Tokyo Institute of Technology, 2-12-1 Ookayama, Meguro, Tokyo 152-8551, Japan}
\author{K. Muraki}
\affiliation{NTT Basic Research Laboratories, NTT Corporation, 3-1 Morinosato-Wakamiya, Atsugi, Kanagawa 243-0198, Japan}
\author{T. Fujisawa}
\email[]{fujisawa@phys.titech.ac.jp}
\affiliation{Department of Physics, Tokyo Institute of Technology, 2-12-1 Ookayama, Meguro, Tokyo 152-8551, Japan}

\date{\today}

\maketitle

{\bf 
The model of interacting fermion systems in one dimension known as Tomonaga-Luttinger liquid (TLL) \cite{tomonaga:1950,luttinger:1963} provides a simple and exactly solvable theoretical framework, predicting various intriguing physical properties.
Evidence of TLL has been observed as power-law behavior in the electronic transport and momentum-resolved spectroscopy on various types of one-dimensional (1D) conductors \cite{tarucha:1995,bockrathet:1999,grayson:1998}.
However, these measurements, which rely on dc transport involving tunneling processes, cannot identify the eigenmodes of the TLL, {\it i.e.}, collective excitations characterized by non-trivial effective charge $e^{\ast}$ and charge velocity $v^{\ast}$.
The elementary process of charge fractionalization \cite{safi:1995,pham:2000,imura:2002}, a phenomenon predicted to occur at the junction of a TLL and non-interacting leads, has not been observed.
Here we report time-resolved transport measurements \cite{kamata:2010} on an artificial TLL comprised of coupled integer quantum Hall edge channels \cite{berg:2009}, successfully identifying single charge fractionalization processes.
An electron wave packet with charge $e$ incident from a non-interacting region breaks up into several fractionalized charge wave packets at the edges of the artificial TLL region, from which $e^{\ast}$ and $v^{\ast}$ can be directly evaluated.
These results are informative for elucidating the nature of TLLs and low-energy excitations in the edge channels \cite{venkatachalam:2012}.
}

\newpage

\begin{figure}[h]
\begin{center}
\includegraphics[scale=0.50]{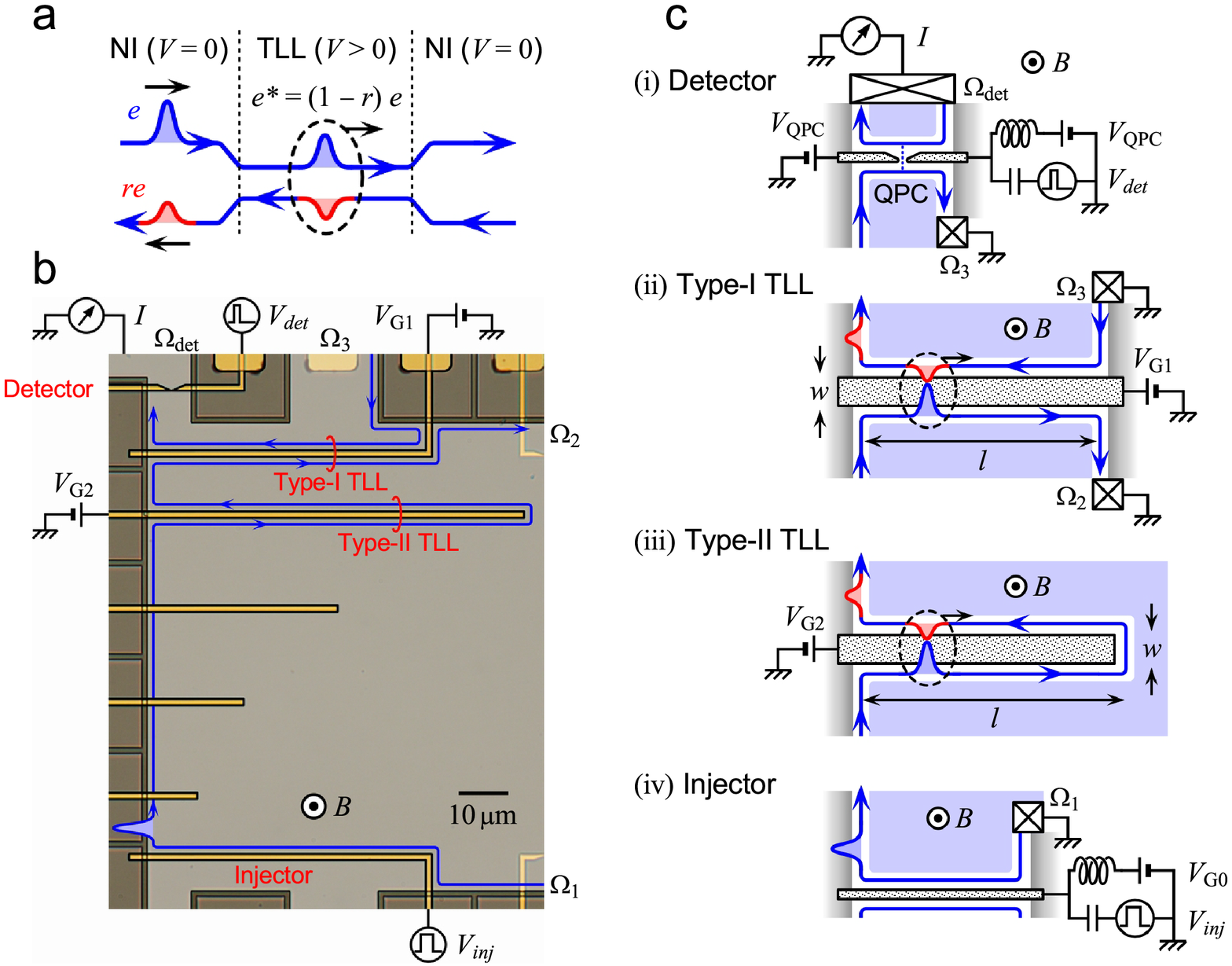}
\caption{{\bf Experimental setup for the time-resolved measurement of the charge fractionalization.}
{\bf a}, 
A schematic illustration of the interacting TLL region with $V > 0$ in between NI regions with $V = 0$.
Injection of charge $e$ from the upper-left lead results in fractionalization into $e^{\ast} = (1 - r) e$ in the TLL region and $r e$ in the lower-left lead.
{\bf b}, 
An optical micrograph of the sample.
Metal gate electrodes (gold regions) are patterned on a 2DES region (light gray region) and etched insulating GaAs regions (dark gray regions).
The 2DES located 90~nm below the surface has a density of $1.45 \times 10^{11}$~cm$^{-2}$ and a low-temperature mobility of $4.0 \times 10^{5}$~cm$^{2}$/Vs.
Chiral 1D edge channels (thick blue lines with arrows) are formed along the edge of the 2DES in a strong perpendicular magnetic field $B$.
{\bf c}, 
Schematic illustrations (not to scale) of the sample consisting (i) a time-resolved charge detector, (ii) Type-I TLL region, (iii) Type-II TLL region, and (iv) a pulsed-charge injector.
The bulk regions (light blue regions) partially screen the unwanted long-range Coulomb interaction.
Excess charge of $Q \sim 150e$, which is roughly estimated from numerical calculation of the electrostatic depletion \cite{larkin:1995}, is injected at the falling edge of a voltage step $V_{inj} = 5$~mV applied to an injection gate in (iv).
The QPC detector in (i) is set at the pinched-off regime, and one of the gate voltage is modulated by a voltage pulse of the height of 0.2~V for the period of 80~ps to temporally control the transmission probability of the QPC.
Average current $I$ through the QPC as a function of the time interval $t$ between two voltage pulses is measured at the detection Ohmic contact $\Omega_{\mathrm{det}}$ under the pulse pattern repeated at 25~MHz.
The gate metal for (ii) Type-I and (iii) Type-II TLL regions have effective length of $l = 68$ and $l = 10 \sim 80$~$\mu$m, respectively, and a width of $w = 1$~$\mu$m.
}
\label{Setup}
\end{center}
\end{figure}

Charge transport in 1D conductors like quantum wires or quantum point contacts (QPCs) are often described by non-interacting electronic modes travelling to the right and left directions \cite{wees:1988}.
Such a single-particle picture breaks down in the presence of Coulomb interaction, where the system is described by the non-chiral TLL model \cite{tomonaga:1950,luttinger:1963} and its elementary excitation as collective modes of charge density waves, historically referred to as `plasmons' \cite{kane:1992}.
The Coulomb interaction enhances the charge velocity in each mode from the Fermi velocity to $U$, where $U$ is often used as a measure of the intra-mode interaction.
When the counter-propagating modes are coupled to form TLL plasmon modes with a coupling ratio of $r=V/2U$, the charge velocity is reduced to $v^{\ast}=\sqrt{U^{2}-V^{2}}$ by the inter-mode interaction $V$ \cite{chang:2003}.
Here we consider a system consisting of two counter-propagating modes (Fig.~\ref{Setup}a), which are interacting in the middle (TLL region with $V>0$) but not interacting outside the middle region (non-interacting (NI) regions with $V=0$).
In the interacting region, a transport eigenmode involves charge $e$ in the right-moving mode dragging a small amount of charge $-re$ in the left-moving mode, constituting non-trivial effective charge $e^{\ast}=(1-r)e$, which induces various TLL phenomena.
However, previous studies searching for such transport eigenmodes have resulted in normal quantized conductance ($e^{2}/h$) or a power-law divergence associated with tunneling or scattering of electrons \cite{tarucha:1995,bockrathet:1999}.
Elementary excitation of the charge $e^{\ast}$ is expected in the charge fractionalization process \cite{safi:1995,pham:2000,imura:2002} that arises at the junction of NI region ($V=0$) and the TLL region ($V>0$).
When a charge $e$ is injected from a NI mode, only a fraction $e^{\ast}$ is induced in the TLL region and the rest ($e-e^{\ast}=re$) should be reflected back in the other NI mode.
However, conventional dc measurement cannot detect the fractionalization since similar multiple fractionalizations at the right and left junction completely cancel all traces.
We successfully observed such multiple fractionalization processes by means of time-resolved experiments.

We used an artificial TLL consisting of two counter-propagating quantum Hall edge channels \cite{berg:2009}.
Fig.~\ref{Setup}b shows the sample patterned on a GaAs/AlGaAs heterostructure with chiral 1D edge channels formed along the edge of the two-dimensional electronic system (2DES) in a strong perpendicular magnetic field $B$.
We conducted measurements with the bulk filling factor $\nu$ between 1 and 2, where a single edge channel with spin-up electrons in the lowest Landau level is formed.
The bulk regions with spin-down electrons partially screen the unwanted long-range Coulomb interaction.
As shown in Figs.~\ref{Setup}c~(ii) and (iii), artificial TLL can be formed in a pair of counter-propagating edge channels along both sides of a narrow gate metal.
The two channels are designed to have sufficient Coulomb interaction $V$ but negligible tunneling effect for a sufficiently large negative gate voltage.
Outside the TLL regions, isolated channels formed along the etched regions work effectively as NI chiral leads for injecting and extracting charges from the TLL regions.
Two types of TLL regions are investigated; type-I with NI leads on both ends (Fig.~\ref{Setup}c~(ii)), and type-II with NI leads only on the left and a closed end on the right (Fig.~\ref{Setup}c~(iii)).
We can selectively activate one of the TLL regions by applying appropriate voltages ($V_{\mathrm{G1}}$ and $V_{\mathrm{G2}}$).

We employed time-resolved charge transport measurement by using high-speed voltage pulses \cite{ashoori:1992,kamata:2010}.
As shown in Fig.~\ref{Setup}c~(iv), a wave packet with non-equilibrium charge $Q$ is generated in the NI edge channel by applying a voltage step of the height $V_{inj}$ to the injection gate.
The wave packet propagates in the channel as an edge magnetoplasmon (EMP) mode, which is a collective mode with finite $U$ \cite{fetter:1985,volkov:1988,talyanskii:1992}.
The wave packet introduced into the TLL region undergoes charge fractionalization processes \cite{berg:2009}.
The reflected charge travels along the outgoing NI lead toward the QPC detector (Fig.~\ref{Setup}c~(i)) \cite{kamata:2010}.
All measurements were carried out at about 300~mK.

\begin{figure}[h]
\begin{center}
\includegraphics[scale=0.55]{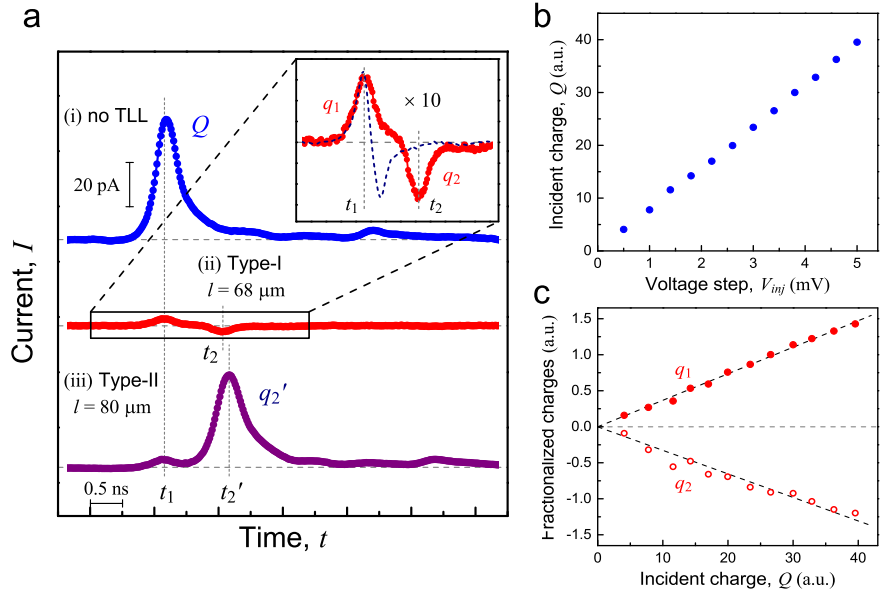}
\caption{{\bf Observed charge wave packets injected and extracted from the TLL regions.}
{\bf a}, 
Charge waveforms observed when (i) no TLL regions are activated, (ii) Type-I TLL region is activated, and (iii) Type-II TLL region is activated.
Traces are vertically offset for clarity.
The inset shows the magnified waveform for (ii).
The derivative of the trace (i) is shown as a dashed curve, which does not agree with our result.
{\bf b},
The area $Q$ of the wave packet in the trace (i) of {\bf a}, plotted as a function of $V_{inj}$.
{\bf c},
The area $q_{1}$ and $q_{2}$ of the first and second wave packets, respectively, in the trace (ii) of {\bf a}, plotted as a function of $Q$.
Dashed lines are results of linear fitting.
}
\label{Waveform}
\end{center}
\end{figure}

Typical waveforms at $\nu=1.5$ ($B=4.0$~T) are summarized in Fig.~\ref{Waveform}a.
The top trace~(i) represents the incident wave packet measured when no TLL regions are activated with $V_{\mathrm{G1}}=V_{\mathrm{G2}}=0$~V.
The traces~(ii) and (iii) show multiple wave packets reflected from Type-I and Type-II TLL regions, respectively (the magnified trace for (ii) is shown in the inset).
Both traces have a small positive wave packet at $t=t_{1}$, which is identical to the arrival time of the incident wave packet to the TLL regions.
The first fractionalization process at the left junction manifests itself at this timing $t_{1}$.
TLL theory predicts its charge to be $q_{1}=rQ$, and our data suggests $r\sim0.04$.
The second wave packet shows a small negative peak at ${t=t_{2}}$ in the trace~(ii), and a large positive peak at $t=t_{2}^{\prime}$ in the trace~(iii).
The charge in the second wave packet depends on the nature of the right end of the TLL regions.
In the case of Type-I TLL region, fractionalization at the right junction and subsequent fractionalization at the left junction should yield the charge $q_{2}=-r(1-r^{2})Q$ in the second wave packet.
This is consistent with our observation of more or less equal magnitude with opposite sign of the first and second wave packets in the trace~(ii).
As for Type-II TLL region, effective charge $(1-r)Q$ is fully reflected at the right end, which yields the charge $q_{2}^{\prime}=(1-r^{2})Q$ in the second wave packet.
This is also consistent with the observation of the large second wave packet in the trace~(iii).
In this way, the observed wave packets are assigned to charge fractionalization at the both ends of the TLL region.

In order to make our analysis quantitative, the amount of charge in each wave packet is estimated from the area of the waveform.
Figure~\ref{Waveform}b shows the linearity of $Q$ with respect to $V_{inj}$.
The fractionalized charges $q_{1}$ and $q_{2}$ obtained for Type-I TLL region are proportional to $Q$ (Fig.~\ref{Waveform}c).
The linearity ensures Ohmic connection between the NI leads and the TLL regions.
The fractionalization ratio $r\sim0.04$, which is related to the inter-mode interaction parameter $g_{2}$ in the TLL as $g_{2}=(1-r)/(1+r)\sim0.92$ \cite{safi:1995,berg:2009}, and hence, effective charge $e^{\ast}=(1-r)e\sim0.96e$ can be extracted from the slope of $q_{1}$ vs $Q$.
Moreover, by measuring the time interval between the two wave packets, which can be regarded as the time of flight for the round trip of the TLL region with length $l$, the effective charge velocity $v^{\ast}\sim150$~km/s can be obtained when corresponding gate voltages are set at $-0.20$~V.

\begin{figure}[h]
\begin{center}
\includegraphics[scale=0.60]{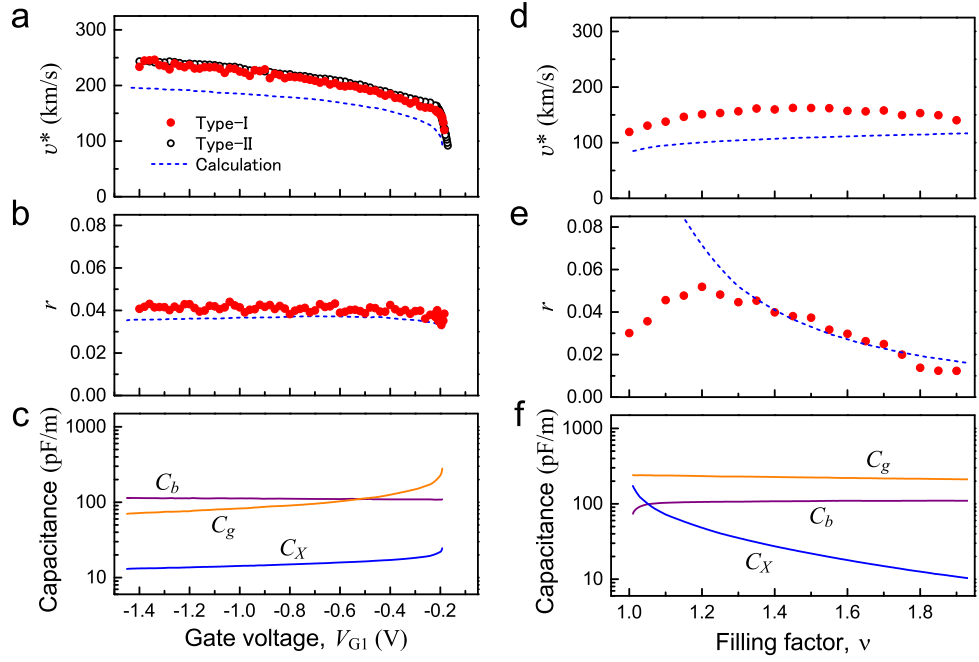}
\caption{{\bf Gate voltage and filling factor dependence of the TLL parameters.}
{\bf a}, {\bf b} and {\bf c},
Gate voltage $V_{\mathrm{G1}}$ dependence of the charge velocity $v^{\ast}$ in {\bf a}, the fractionalization ratio $r$ in {\bf b}, and coupling capacitances in {\bf c}.
Dashed lines in {\bf a} and {\bf b} are obtained from the calculated capacitances in {\bf c} without any fitting parameters.
{\bf d}, {\bf e} and {\bf f},
Filling factor $\nu$ dependence of $v^{\ast}$ in {\bf d}, $r$ in {\bf e}, and coupling capacitances in {\bf f}.
Dashed lines in {\bf d} and {\bf e} are obtained from the calculated capacitances in {\bf f} without any fitting parameters.
}
\label{TLLparameter}
\end{center}
\end{figure}

\begin{figure}[h]
\begin{center}
\includegraphics[scale=0.60]{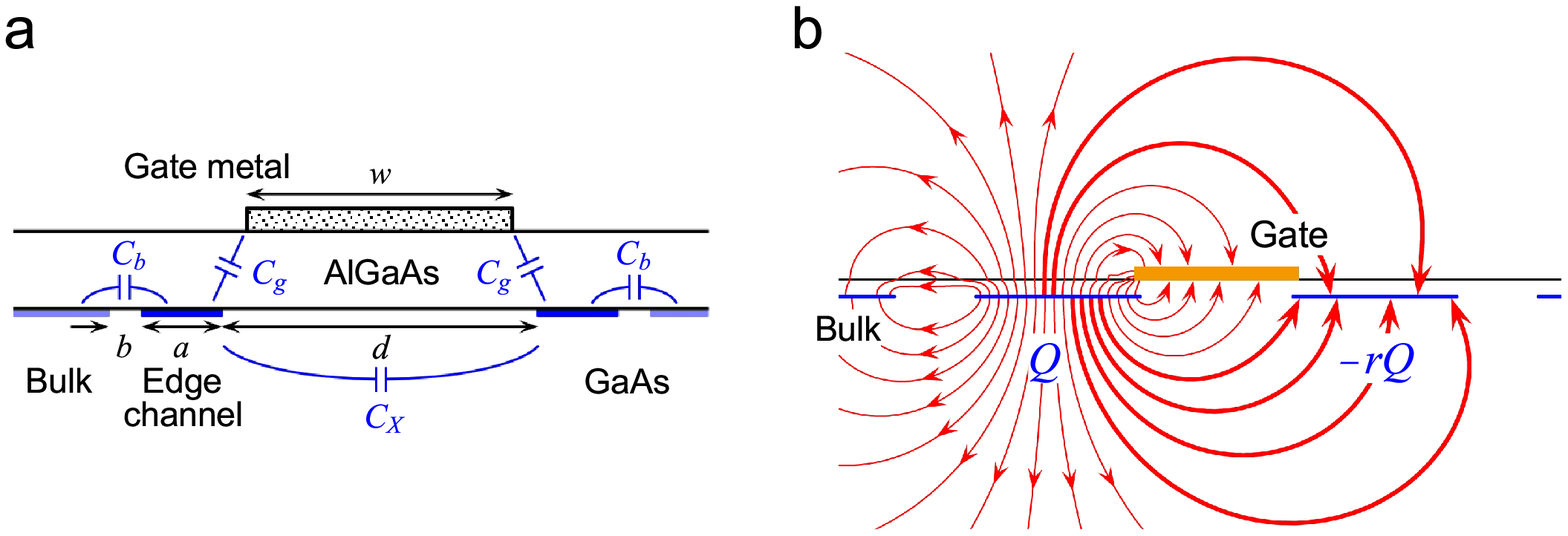}
\caption{{\bf Microscopic model of Coulomb interaction with electrostatic capacitances.}
{\bf a},
Schematic cross section (not to scale) of the artificial TLL region with coupling capacitors, $C_{g}$, $C_{b}$, and $C_{X}$.
{\bf b},
Lines of electric force (arrows) associated with the TLL mode with $Q$ in the left and $- r Q$ in the right channel.
Most of the lines goes to the gate and bulk regions, while a few lines (thick arrows) connect the two channels with representing $C_{X}$.
Calculations were made with the parameters ($a = 1$~$\mu$m, $b = 0.5$~$\mu$m, $w = 1$~$\mu$m) for $\nu = 1.5$ and $V_{\mathrm{G}} = - 0.20$~V.
}
\label{CModel}
\end{center}
\end{figure}

We investigated how the parameters in the artificial TLL change with the gate voltage $V_{\mathrm{G1}}$ (Figs.~\ref{TLLparameter}a and \ref{TLLparameter}b), and the filling factor $\nu$ (Figs.~\ref{TLLparameter}d and \ref{TLLparameter}e).
In our artificial TLL, $v^{\ast}$ and $r$ strongly depend on $V_{\mathrm{G1}}$ and $\nu$, respectively.
These variations can be reproduced by considering the Coulomb interactions characterized with the coupling capacitances between the edge channels, $C_{X}$, as well as between the channel and the gate metal, $C_{g}$, and the bulk region, $C_{b}$, as depicted in Fig.~\ref{CModel}a.
Capacitances obtained from an electrostatic analysis (see Methods) are plotted as functions of $V_{\mathrm{G1}}$ and $\nu$ in Figs.~\ref{TLLparameter}c and \ref{TLLparameter}f, respectively.
We find the TLL parameters represented by the capacitances as
\begin{equation}
v^{\ast}=\sqrt{U^2-V^2}
\sim\frac{\sigma_{xy}}{\sqrt{C_{ch}(C_{ch}+2C_{X})}},
\label{TLLv}
\end{equation}
\begin{equation}
r=\frac{V}{2U}\sim\frac{C_{X}}{2C_{ch}},
\label{TLLr}
\end{equation}
where $C_{ch}=C_{g}+C_{b}$ is used (see Supplementary Information).
The good agreements between the calculated values (dashed lines in Figs.~\ref{TLLparameter}a, \ref{TLLparameter}b, \ref{TLLparameter}d, and \ref{TLLparameter}e) and the experimental values supports the validity of our analyses and implies that TLL parameters can be designed with the channel structures.

The dependence of the parameters on $V_{\mathrm{G1}}$ and $\nu$ can be understood in terms of screening effects.
Figure~\ref{CModel}b shows a cross-sectional view of typical electric forces (arrows) associated with the transport eigenmode of the TLL.
Electrostatic solution indicates that a few lines (thick arrows) of electric force are connected between the edge channels through the GaAs layer while others (thin arrows) are screened by the gate and bulk regions.
$V_{\mathrm{G1}}$ determines the distance between the edge channel and the gate metal, which significantly modifies $C_{g}$ and $C_{X}$ (Fig.~\ref{TLLparameter}c).
This changes the channel capacitance $C_{ch}(=C_{g}+C_{b})$ and $v^{\ast}$, leaving $r$ unchanged (Fig.~\ref{TLLparameter}b).
On the other hand, $\nu$ changes the widths of the compressible and incompressible regions ($a$ and $b$) \cite{chklovskii:1992,kumada:2011}, which mainly influence $C_{X}$ and $r$, leaving $C_{g}$ and $C_{b}$ less affected (Figs.~\ref{TLLparameter}e and \ref{TLLparameter}f).
In this way, $v^{\ast}$ and $r$ can be tuned separately with $V_{\mathrm{G1}}$ and $\nu$, respectively.

In summary, charge fractionalization at the edges of an artificial TLL region is identified with our time-resolved measurement of charge wave packets.
Obtained parameters are consistent with the TLL model as well as microscopic analysis of Coulomb interaction with electrostatic environment.
Our measurement scheme and analysis would provide in-depth investigation of interacting electrons in various TLLs as well as quantum Hall edge channels.
More generally, the strength of inter-channel interaction $V$ changes as a function $V(x)$ of the position $x$ along the channel.
Basically, if the time-resolution is sufficiently high, the measurement could provide time-domain reflectometry of $V(x)$, as the reflection is sensitive to its derivative $dV(x)/dx$.
The time-domain reflectometry would elucidate various phenomena, such as spin-charge separation in spinfull TLLs \cite{lee:1997,bocquillon:2013}, chargeless heat transport in integer and fractional quantum Hall regime \cite{venkatachalam:2012}, and dissipation processes in coherent electron transport \cite{roulleau:2008}.

\appendix
\section{Methods}
\noindent{\bf Calculation for the coupling capacitances.}
As shown in the cross section of the TLL region in Fig.~\ref{CModel}a, the interactions between the conductive regions (gate metal, bulk regions, and edge channels) can be characterized by the coupling capacitances ($C_{g}$, $C_{b}$, and $C_{X}$).
We numerically calculated these capacitances by solving two-dimensional Laplace equation with certain boundary conditions.
For simplicity, all conductive regions are assumed to be ideal conductors.
The width of the edge channel (compressible region), $a$, and the incompressible gap, $b$, are predetermined from an analytical solution for a similar structure \cite{chklovskii:1992}, and the distance between the edge channels, $d$, is also determined from an electrostatic solution \cite{larkin:1995}.

\section{Acknowledgments}
We are grateful to K.-I.~Imura and M.~Nakamura for fruitful discussions and M.~Ueki for experimental support.
This work was partially supported by Grants-in-Aid for Scientific Research (21000004, 11J09248) and the Global Center of Excellence Program from the MEXT of Japan through the ``Nanoscience and Quantum Physics" Project of the Tokyo Institute of Technology.

\section{Author contributions}
H.K. performed the experiments, analysed the data and wrote the manuscript.
T.F. supervised the research.
K.M. grew the wafer.
All authors discussed the results and commented on the manuscript.

\section{Competing financial interests}
The authors declare no competing financial interests.

\clearpage

\section{Supplementary information}
\noindent{\bf 1. Derivation of equations (1) and (2) in the main manuscript}

Wave equation for charge density wave in the form of an edge magnetoplasmon mode can be found in many articles \cite{fetter:1985,aleiner:1994,zhitenev:1995}.
Here, we use a simplified yet general form with a channel capacitance $C_{ch}$ defined per unit length, which relates the potential $V(x,t)$ and excess charge density $\rho(x,t)$ at position $x$ along the channel and time $t$ as $\rho(x,t)=C_{ch}V(x,t)$.
Associated `Hall' current $I(x,t)=\sigma_{xy}V(x,t)$ along the channel changes the charge distribution through the continuity equation $\partial \rho(x,t)/\partial t=-\partial I(x,t)/\partial x$.
Considering the conductance of the channel $I(x,t)/V(x,t)=\sigma_{xy}$, the unidirectional wave equation reads
\begin{equation}
\frac{\partial}{\partial t}\rho(x,t)
=-\frac{\sigma_{xy}}{C_{ch}}
\frac{\partial}{\partial x} \rho(x,t),
\end{equation}
where $\sigma_{xy}/C_{ch}$ gives the charge velocity in uncoupled channels.

Coupled charge density wave in our artificial TLL can be considered in a pair of two counter-propagating edge channels $i$ ($i=1$ for propagating to the positive direction; $i=2$ for negative direction).
They are coupled with the cross capacitance $C_{X}$ per unit length.
The relation between the charge density $\rho_{i}(x,t)$ and the potential $V_{i}(x,t)$ can be described in a matrix form as
\begin{equation}
\left(
\begin{array}{c}
\rho_{1} \\
\rho_{2}
\end{array}
\right)
=
\left(
\begin{array}{cc}
C_{ch}+C_{X} & -C_{X} \\
-C_{X} & C_{ch}+C_{X}
\end{array}
\right)
\left(
\begin{array}{c}
V_{1} \\
V_{2}
\end{array}
\right).
\end{equation}
Using the continuity equation, the coupled wave equation for $\rho_{i}(x,t)$ is given by
\begin{equation}
\frac{\partial}{\partial t} \left(
\begin{array}{c}
\rho_{1} \\
\rho_{2}
\end{array}
\right)
=-\frac{\partial}{\partial x}\left(
\begin{array}{cc}
U & V \\
-V & -U
\end{array}
\right)
\left(
\begin{array}{c}
\rho_{1} \\
\rho_{2}
\end{array}
\right),
\label{S3}
\end{equation}
where $U$ and $V$ are intra- and inter-channel interaction strengths, respectively.
They are given by 
\begin{equation}
U=\frac{\sigma_{xy}}{C_{ch}}
\frac{C_{ch}+C_{X}}{C_{ch}+2C_{X}},
\end{equation}
\begin{equation}
V=\frac{\sigma_{xy}}{C_{ch}}
\frac{C_{X}}{C_{ch}+2C_{X}}.
\end{equation}
The wave equation (\ref{S3}) exhibits transport eigenmode $-\rho_{2}/\rho_{1}$ or $-\rho_{1}/\rho_{2}=V/(U+\sqrt{U^{2}+V^{2}})$, which is defined as the coupling factor $r$, and the velocity $v^{\ast}=\sqrt{U^{2}-V^{2}}$.
Inserting the channel capacitance $C_{ch}=C_{g}+C_{b}$ for the model defined in the main manuscript gives equations (1) and (2).

\newpage

\noindent{\bf 2. Reflection coefficient}
\begin{figure}[h]
\begin{center}
\includegraphics[scale=0.68]{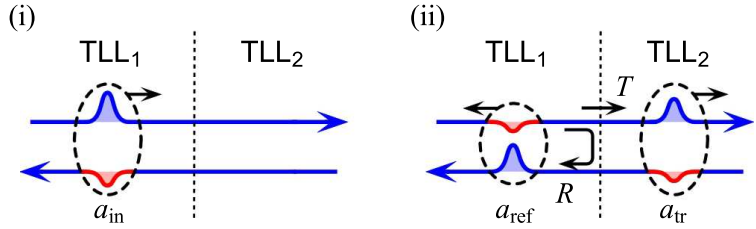}
\caption{
{\bf Reflection coefficient between two TLL regions.}
Schematic illustration of (i) the incident right moving mode $a_{\mathrm{in}} \mathbf{u}_{\mathrm{R1}}$ in TLL$_{1}$, (ii) the reflected left moving mode $a_{\mathrm{ref}}\mathbf{u}_{\mathrm{L1}}$ in TLL$_{1}$ and the transmitted right moving mode $a_{\mathrm{tr}}\mathbf{u}_{\mathrm{R2}}$ in TLL$_{2}$.}
\label{TLL12}
\end{center}
\end{figure}

Consider two TLL regions TLL$_{1}$ and TLL$_{2}$ with the coupling coefficient $r_{1}$ and $r_{2}$, respectively, connected with each other as shown in Fig.~\ref{TLL12}.
We shall describe the eigenmodes in TLL$_{i}$ as $\mathbf{u}_{\mathrm{R}i}=(1,-r_{i})$ for right moving mode and $\mathbf{u}_{\mathrm{L}i}=(-r_{i},1)$ for left moving mode.
Incident wave of amplitude $a_{\mathrm{in}}$ from TLL$_{1}$ is partially transmitted into the TLL$_{2}$ with the amplitude $a_{\mathrm{tr}}$ and reflected back to TLL$_{1}$ with the amplitude $a_{\mathrm{ref}}$.
Charge conservation law determines a relation
\begin{equation}
a_{\mathrm{in}}
\left(
\begin{array}{c}
1 \\
-r_{1}
\end{array}
\right)
=a_{\mathrm{ref}}
\left(
\begin{array}{c}
-r_{1} \\
1
\end{array}
\right)
+a_{\mathrm{tr}}
\left(
\begin{array}{c}
1 \\
-r_{2}
\end{array}
\right),
\end{equation}
from which we obtain the transmission coefficient $T\equiv a_{\mathrm{tr}}/a_{\mathrm{in}}=(1-r_{1}^{2})/(1-r_{1}r_{2})$ and the reflection coefficient $R\equiv a_{\mathrm{ref}}/a_{\mathrm{in}}=(r_{2}-r_{1})/(1-r_{1}r_{2})$.
For the abrupt junction between non-interacting channel ($r_{1}=0$) and TLL region ($r_{2}$) defined in Fig.~\ref{Setup}a in the main manuscript, we used $T=1$ and $R=r_{2}$ to evaluate the reflected charge.
The above formula indicates that the reflection is associated with the difference of the coupling coefficients, and can be used to develop a time-domain reflectometry.

\newpage

\noindent{\bf 3. $V_{\mathrm{G1}}$ and $\nu$ dependence of charge waveforms}
\begin{figure}[h]
\begin{center}
\includegraphics[scale=0.65]{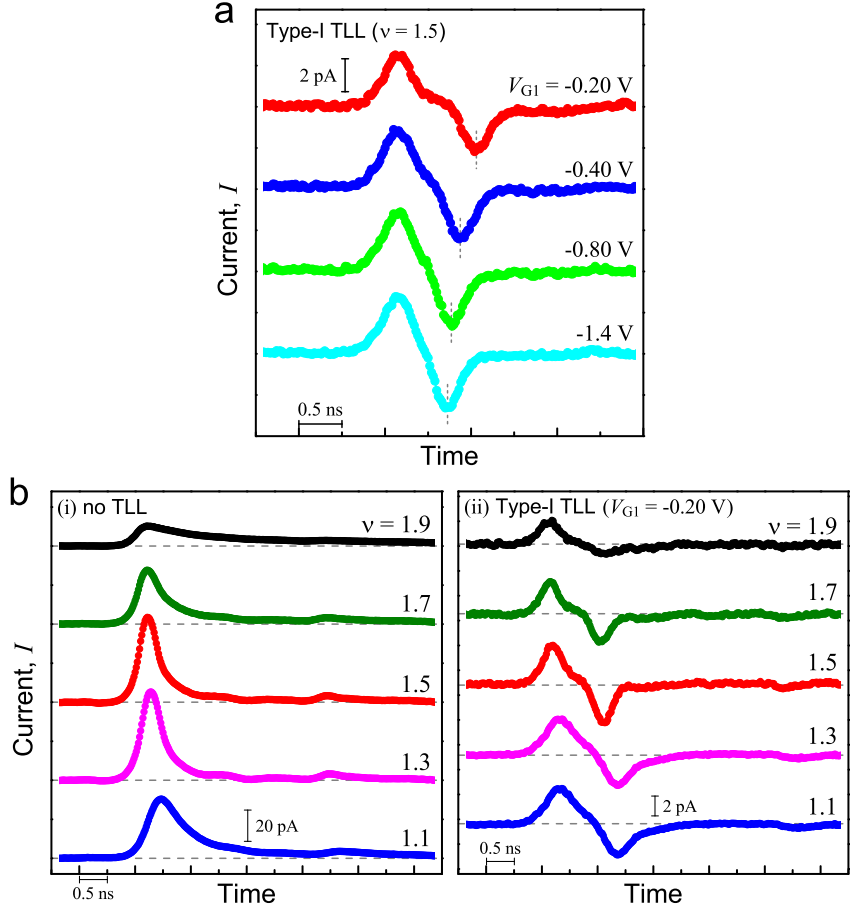}
\caption{
{\bf $V_{\mathrm{G1}}$ and $\nu$ dependence of charge waveforms.}
{\bf a},
$V_{\mathrm{G1}}$ dependence of charge waveforms for the Type-I TLL region observed when $\nu=1.5$.
Traces are vertically offset for clarity.
{\bf b},
$\nu$ dependence of charge waveforms observed when (i) no TLL regions are activated, (ii) Type-I TLL region is activated with the gate voltage $V_{\mathrm{G1}}=-0.20$~V.
Traces are vertically offset for clarity.
}
\label{waveforms}
\end{center}
\end{figure}

\noindent{\bf 4. Charge waveforms}

It is important to distinguish real signals from spurious ones in Fig.~\ref{Waveform}a in the main manuscript.
Indeed, the additional small ripples seen at $t\sim t_{1}+3.5$~ns and $t\sim t_{2}^{\prime}+3.5$~ns in the traces~(i) and (iii) are due to ripples in our voltage pulse waveform and irrelevant to the fractionalization processes.
Electrostatic crosstalk between the leads also induces spurious signals in this type of experiments \cite{hashisaka:2011}.
However, such crosstalk can be ruled out, as it should appear as a derivative of the incident wave (shown as a dashed line in the inset), which cannot explain our observations.
Although it is also known that plasmon mode exists in the bulk compressible region, such bulk plasmon mode is dissipative and exhibits a lower velocity \cite{volkov:1988}.
Therefore, we do not consider the effects from the bulk plasmon mode in the analysis carried out in the main manuscript.
However, we find significant broadening of the peaks for $\nu$ less than 1.3 (Figs.~\ref{waveforms}b and \ref{waveforms}c).
This could be associated with localized bulk regions with fewer electrons.
The scattering of charge density waves by the randomly-distributed localized regions may broaden the waveform \cite{guven:2003,kumada:2011}.
Another possible reason for the broadening could be the reduced screening effect for long-range Coulomb interaction at smaller $\nu$.
Our previous results suggest long-range interaction over 100~$\mu$m near integer filling factors \cite{hashisaka:2012}.
In practice, the change in the inter-channel interaction $V$ is not as abrupt as illustrated in Fig.~\ref{Setup}a in the main manuscript, but smooth in a finite transition length.
The observed width ($\sim0.4$~ns) in the reflected wave (ii) of Fig.~\ref{Waveform}a in the main manuscript indicates that the transition length is less than 70~$\mu$m (the upper bound ignoring other reasons), by assuming identical $U$ in the leads and the TLL region.
\\

\noindent{\bf 5. Time-resolved measurement vs dc measurement}

In Type-I TLL region (Fig.~\ref{Setup}a~(ii) in the main manuscript) the NI leads connected to the injection and detection circuits are electrically isolated, so that current should always be zero in dc transport experiments.
In our case (the trace~(ii) of Fig.~\ref{Waveform}a in the main manuscript), sum of the charges of all reflected wave packets including higher-order reflection expected at $t>t_{2}$ should be zero.
Since $r$ is small in our device, the higher-order terms are always below the detection limit, and therefore sum of the two wave packets is almost zero within the noise level of our measurement.
As for Type-II TLL region (Fig.~\ref{Setup}a~(iii) in the main manuscript), sum of all charges should be and actually is identical to the incident charge.
In this way, multiple fractionalization can only be resolved in time-resolved measurement but not in dc measurements.
Another notable feature of our time- resolved measurement is that it does not involve tunneling processes.
If the NI leads and the TLL regions are connected through a tunneling barrier, only a fixed charge of $e$ rather than arbitrary charge ($rQ$) can be reflected from the TLL region, where individual fractionalization processes cannot be resolved with tunneling experiments.

\end{document}